\documentclass[preprint,showpacs]{revtex4}
\usepackage{latexsym}

\begin{document}
\preprint{\vbox{\hbox{\tt gr-qc/0607127}}}
\title{Remarks on the Wave Function of the Universe in the Dilaton Cosmology}
\author{Wontae \surname{Kim}}\email{wtkim@sogang.ac.kr}
\affiliation{Department of Physics, Center for Quantum Spacetime,
 and Basic Science Research Institute,
 Sogang University, C.P.O. Box 1142, Seoul 100-611, Korea}
\author{Edwin J. \surname{Son}}\email{ejson@yonsei.ac.kr}
\affiliation{Institute of Physics and Applied Physics
 and Natural Science Research Institute,
 Yonsei University, Seoul 120-749, Korea}
\author{Myung Seok \surname{Yoon}}\email{yoonms@sejong.ac.kr}
\affiliation{Department of Physics, Sejong University, Seoul 143-747, Korea}

\date{\today}
\bigskip
\begin{abstract}
Motivated by previous works, we study semi-classical cosmological
solutions and the wave function of the Wheeler-DeWitt equation in the
Bose-Parker-Peleg model. We obtain the wave function of the universe
satisfying the suitable boundary condition of the redefined fields,
which has not been considered in previous works. For some limiting
cases, the Wheeler-DeWitt equation is reduced to the Liouville
equation with a boundary, and its solution can be described by
well-known functions. The consistent requirement of the boundary
condition is related to the avoidance of the curvature singularity.
\end{abstract}
\maketitle

In some gravitational systems, one of the most intriguing subjects is
to investigate the universe; however, it is difficult to obtain a
cosmological solution and a wave function of the universe in the
(3+1)-dimensional quantum region, because the quantum gravity is not
yet known in that dimension. For this reason, it has been of interest
to consider renormalizable two-dimensional dilation gravity models as
exactly solvable toy models in which some of the difficulties of the
realistic problem are not present, and these models have been shown to
be useful in investigating various aspects of  Hawking radiation and
black-hole evaporation~\cite{cghs,rst,bpp}. By virtue of these models,
two-dimensional cosmological solutions, which have been used to study
the graceful exit problem and quantum cosmology, were obtained by
various authors~\cite{mr,Rey,gv,Bose,bk,ky}. Especially, in
Ref.~\cite{mr} based on the Russo-Susskind-Thorlacius(RST)
model~\cite{rst}, the Wheeler-DeWitt(WD) equation was considered, and
it was shown that the semi-classical solution could be deduced from
its oscillatory solution, which was obtained by neglecting the
boundary condition of the fields in the Liouville theory. 

On the other hand, there have been cosmological models based on the
Bose-Parker-Peleg(BPP) model~\cite{bpp}, in which it is possible to
obtain the exact metric solution explicitly~\cite{Bose,bk}. Thus, in
this brief report, we would like to obtain the semi-classical
cosmological solutions and the WD equation in this model and study the
wave function of the universe by taking into account the boundary
condition of the involved fields(for WD equation and the wave
functions in four or higher dimensions, see Refs.~\cite{imz}). For
some special limits, the WD equation is reduced to the Liouville
equation with a boundary, and its solution is described by using
modified Bessel functions. This boundary condition enables us to
remove the cosmological singularity. 

We begin with the dilaton gravity action coupled to $N$-conformal
matter fields $f_i$, including the one-loop Polyakov action with local
covariant terms as in the BPP model~\cite{bpp}:
\begin{eqnarray}
S &=& S_0 + S_\mathrm{qt} \label{eq:action} \\
S_0 &=& \frac{1}{2\pi} \int d^2x \sqrt{-g} \left\{ e^{-2\phi} \left[
  R + 4(\nabla \phi)^2 + 4\lambda^2 \right] - \frac12
  \sum_{i=1}^{N} (\nabla f_i)^2 \right\}, \label{eq:action:cl} \\
S_\mathrm{qt} &=& \frac{1}{2\pi} \int d^2x \sqrt{-g} \left[
  -\frac{\kappa}{4} R \frac{1}{\Box} R + \kappa (\nabla \phi)^2 -
  \kappa \phi R \right], \label{eq:action:qt}
% S &=& \int d^2x \left[ \mathcal{L}_\mathrm{DG} +
%   \mathcal{L}_\mathrm{f} + \mathcal{L}_\mathrm{qt} \right],
%   \label{eq:action} \\
% \mathcal{L}_\mathrm{DG} &=& \frac{1}{2\pi} \sqrt{-g} e^{-2\phi} \left[
%   R + 4(\nabla \phi)^2 + 4\lambda^2 \right], \label{eq:action:DG} \\
% \mathcal{L}_\mathrm{f} &=& \frac{1}{2\pi} \sqrt{-g} \left[ - \frac12
%   \sum_{i=1}^{N} (\nabla f_i)^2 \right], \label{eq:action:f} \\
% \mathcal{L}_\mathrm{qt} &=& \frac{1}{2\pi} \sqrt{-g} \left[
%   -\frac{\kappa}{4} R \frac{1}{\Box} R + \kappa (\nabla \phi)^2 -
%   \kappa \phi R \right], \label{eq:action:qt}
\end{eqnarray}
where $\kappa = N\hbar/12$.
In the conformal gauge, $ds^2=-e^{2\rho}d\sigma^+d\sigma^-$ in the
light-cone coordinates of $\sigma^\pm=t\pm\sigma$, and the total
action in Eq.~(\ref{eq:action}) can be written as 
\begin{widetext}
\begin{equation}
S = \frac1\pi \int d^2x \left[ -\frac1\kappa \partial_+ \chi
  \partial_- \chi + \frac1\kappa \partial_+ \Omega \partial_- \Omega +
  \lambda^2 e^{2(\chi-\Omega)/\kappa} + \frac12 \sum_i
  \partial_+ f_i \partial_- f_i \right] \label{eq:action:eff}
\end{equation}
\end{widetext}
in terms of new fields $\chi$ and $\Omega$ defined by
$\chi=e^{-2\phi}+\kappa(\rho-\phi)$ and $\Omega=e^{-2\phi}$. Note that
the effective coupling is simply defined as
$g_\mathrm{eff}^2=e^{-2\phi}$ in this model. Then, the equations of
motion and the constraints are given by
\begin{eqnarray}
& & \partial_+ \partial_- \chi = \partial_+ \partial_- \Omega =
  -\lambda^2 e^{2(\chi-\Omega)/\kappa},  \label{eq:eom:Omega} \\
& & \partial_+ \partial_- f_i = 0,  \label{eq:eom:matter} \\
& & \kappa t_\pm(\sigma^\pm) = -\frac1\kappa \partial_\pm \chi
  \partial_\pm \chi + \partial_\pm^2 \chi + \frac1\kappa \partial_\pm
  \Omega \partial_\pm \Omega \nonumber \\
& & \qquad \qquad \quad
 + \frac12 \sum_i  \partial_\pm f_i
  \partial_\pm f_i,  \label{eq:constraints}
\end{eqnarray}
where the integration functions $t_\pm(\sigma^\pm)$ reflect the
non-locality of the quantum anomaly term and are to be determined by
appropriate boundary conditions. It has been pointed out in
Ref.~\cite{bpp} that there is no Minkowski vacuum solution in the BPP
model; two possible solutions are an asymptotic Minkowski vacuum and a
flat spacetime in thermal equilibrium. 

Following the cosmological principle of a homogeneous spacetime, we
set the matter distribution to $f_i=f_i(t)$, and we choose the
coordinate $t$ as $\chi-\Omega=\kappa(\rho-\phi)=\kappa\lambda t$ from
Eq.~(\ref{eq:eom:Omega}). As in Ref.~\cite{mr}, the free-field
solution $\rho -\phi =0$ is possible, which gives a different
solution. This situation is more or less different from the case of a
black hole in that any choice of the coordinate system gives the same
black-hole solution apart from a shift of the event horizon. Now, the
solutions are obtained as
\begin{eqnarray}
\Omega &=& -e^{\lambda(\sigma^+ + \sigma^-)} + C(\sigma^+ +
  \sigma^-) + M/\lambda, \label{eq:sol} \\
f_i &=& \frac{p_i}{2}(\sigma^+ + \sigma^-) + b_i,
  \label{eq:sol:matter}
\end{eqnarray}
where $C$, $M$, $p_i$, and $b_i$ are constants. Here, if an asymptotic
Minkowski vacuum is assumed, then $\rho$ and its derivatives, and the
expectation value of the energy-momentum tensors $\langle T_{\pm\pm}^f
\rangle = \kappa \left[ \partial_\pm^2 \rho - (\partial_\pm\rho)^2 -
  t_\pm(\sigma^\pm) \right]$ vanish on the boundary, which results in
$t_\pm=0$. Thus, the constant $C$ in the general solution in
Eq.~(\ref{eq:sol}) is fixed from the constraints in
Eq.~(\ref{eq:constraints}) as
\begin{equation}
\Omega = e^{-2\phi} = -e^{\lambda(\sigma^+ + \sigma^-)} +
  \lambda\mu^2 (\sigma^+ + \sigma^-) + \frac{M}{\lambda},
  \label{eq:sol:Omega}
\end{equation}
where $\mu^2=m^2-\kappa/4$ and $m^2=(1/8\lambda^2)\sum p_i^2$. Note
that the classical solution is easily obtained by setting $\kappa=0$
in Eq.~(\ref{eq:sol:Omega}):
\begin{equation}
e^{-2\phi_\mathrm{cl}} = -e^{\lambda(\sigma^+ + \sigma^-)} +
\lambda m^2 (\sigma^+ + \sigma^-) + \frac{M}{\lambda}.  \label{eq:sol:cl}
\end{equation}

Note that the solution $\Omega$ must be positive,
$\Omega=e^{-2\phi}>0$, such that the coordinate $t$ should have an
upper-bound $t_b$. In addition, there is a lower-bound $t_a$ if
$\mu^2>0$, where $e^{-2\phi}|_{t=t_a,t_b}=0$. Then, the Ricci scalar
is explicitly calculated as
\begin{equation}
R = 4\lambda^2 + 4\lambda^2 \frac{e^{-2\lambda t} (e^{2\lambda t} -
  \mu^2)^2}{-e^{2\lambda t} + 2\lambda\mu^2t + M/\lambda},  \label{eq:R:scalar}
\end{equation}
and it diverges at $t=t_a,t_b$. We hereafter assume $\mu^2>0$ for
simplicity. 

Following the procedure in Refs.~\cite{mr} and~\cite{Polchinski}, we
now quantize our model. Assuming that the spatial coordinate $\sigma$
is periodic, we can expand the fields and their derivatives as
\begin{eqnarray}
& & \Omega(0,\sigma) = \Omega_+(0,\sigma) + \Omega_-(0,\sigma),
  \label{eq:expand:Omega} \\
& & \Omega_\pm(0,\sigma) = \frac12 \Omega_0 \pm \Omega_0^\pm \sigma -
  i \sum_{n\ne0} \frac1n \Omega_n^\pm e^{\pm in\sigma},
  \label{eq:expand:pm} \\
& & \partial_\pm \Omega(0,\sigma) = \sum_{n=-\infty}^{\infty}
\Omega_n^\pm e^{\pm in\sigma},  \label{eq:expand:derivative}
\end{eqnarray}
and $\chi$ and $f_i$ can be similarly expanded. Then, the commutation
relations are defined by
\begin{eqnarray}
& & [\Omega_0,P_\Omega] = i, \quad [\Omega_n^\pm,\Omega_m^\pm] = -
  \frac{n\kappa}{4} \delta_{n+m},  \label{eq:commutation:Omega} \\
& & [\chi_0,P_\chi] = i, \quad [\chi_n^\pm,\chi_m^\pm] =
  \frac{n\kappa}{4} \delta_{n+m},  \label{eq:commutation:chi} \\
& & [f_{i0},P_{f_i}] = i, \quad [f_{in}^\pm,f_{im}^\pm] = -
  \frac{n}{2} \delta_{n+m},  \label{eq:commutation:matter}
\end{eqnarray}
where we used the relations $\Omega_0^\pm=(\kappa/4)P_\Omega$,
$\chi_0^\pm=-(\kappa/4)P_\chi$, and $f_{i0}^\pm=(1/2)P_{f_i}$. From
the constraints in Eq.~(\ref{eq:constraints}), the Virasoro generators
$L_n^\pm$ in the cylinder can be written as
\begin{widetext}
\begin{equation}
L_n^\pm = \sum_m \left[ \frac2\kappa (\Omega_m^\pm \Omega_{n-m}^\pm -
  \chi_m^\pm \chi_{n-m}^\pm) + \sum_i f_{im}^\pm f_{in-m}^\pm \right]
  + 2in\chi_n^\pm - \frac\kappa2 \delta_n - \frac{\lambda^2}{\pi}
  \int_0^{2\pi} d\sigma e^{\mp in\sigma} e^{2(\chi-\Omega)/\kappa}
  \label{eq:Virasoro}
\end{equation}
in terms of the mode expansions, Eqs.~(\ref{eq:expand:pm}) and
(\ref{eq:expand:derivative}). The physical state condition gives the
non-trivial WD equation
\begin{equation}
\left[ \frac\kappa4 \left( \frac{\partial^2}{\partial\chi_0^2} -
  \frac{\partial^2}{\partial\Omega_0^2} \right) - \frac12 \sum_i
  \frac{\partial^2}{\partial f_{i0}^2} - 4\lambda^2
  e^{2(\chi_0-\Omega_0)/\kappa} - \kappa \right] \Psi = 0  \label{eq:WDW}
\end{equation}
\end{widetext}
in the minisuperspace approximation, which is, in fact, equivalent to
satisfying the functional Hamiltonian constraints
$H\Psi=(L_0^++L_0^-)\Psi=0$, where the vacuum wave function is set to
$\Psi=\Psi_p(f_{i0})\Psi_\alpha(\chi_0,\Omega_0)$, because the other
states are annihilated by the excited modes. By defining new fields as
$\chi_+=\chi_0+\Omega_0$ and $\chi_-=(\chi_0-\Omega_0)/\kappa$, we can
rewrite the WD equation as
\begin{equation}
\left[ \frac{\partial}{\partial\chi_+} \frac{\partial}{\partial\chi_-}
  - \alpha - 4\lambda^2 e^{2\chi_-} \right] \Psi_\alpha=0,  \label{eq:WDW:+-}
\end{equation}
where $\alpha \equiv \frac{N}{12} - \frac12 \sum p_i^2$ which yields a
solution in the form of
\begin{equation}
%\begin{eqnarray}
%& & 
\Psi =  \Psi_p  \Psi_\alpha=\exp(i\sum p_i f_{i0}) %\times \nonumber \\
%& & \qquad \qquad 
\exp i \left[ p_- \chi_+ + p_+ \chi_- -
  \frac{2\lambda^2}{p_-} e^{2\chi_-} \right]  \label{eq:WDW:sol}
%\end{eqnarray}
\end{equation}
with $p_+ p_- = - \alpha$. Note that the classical trajectories
associated with the above wave function of the form $\Psi=e^{iS}$ can
be written as
\begin{eqnarray}
& & \chi_- = -p_- t, \quad f_{i0} = p_i t + b_i, \nonumber \\
& & \chi_+ = -p_+ t - \frac{2\lambda^2}{p_-^2} e^{-2p_- t} +
  \textrm{const.,}  \label{eq:WDW:cl}
\end{eqnarray}
by realizing that $\partial_\pm S=-\dot{\chi}_\mp$ and
$\partial_{f_{i0}}S=\dot{f}_{i0}$, where the dots denote derivatives
with respect to $t$~\cite{mr}. This is the definition of classical time
$t$ in terms of the quantum degrees of
freedom~\cite{Banks,Hartle}. Then, the semi-classical solution in
Eq.~(\ref{eq:sol:Omega}) can be reproduced by substituting $\chi_+ =
2\Omega_0 + \kappa \chi_-$, $p_+=-4\lambda m^2$, and $p_-=-\lambda$
into Eq.~(\ref{eq:WDW:cl}).

On the other hand, in this oscillating solution, as pointed out in
Ref.~\cite{mr}, the boundedness of $\Omega$ has not been taken into
account, although it should have been imposed on the wave function as
a boundary condition. In order to take into account the boundedness of
$\Omega$,  we return to the WD equation, Eq.~(\ref{eq:WDW:+-}),
rewritten as 
\begin{equation}
\left[ \frac\kappa4 \left( \frac{\partial^2}{\partial\chi_0^2} -
  \frac{\partial^2}{\partial\Omega_0^2} \right) - \alpha - 4\lambda^2
  e^{2(\chi_0-\Omega_0)/\kappa} \right] \Psi_\alpha = 0  \label{eq:WDW:chiOmega}
\end{equation}
instead of the above light-cone representation. Then, three limiting
cases can be considered: (i) $\chi_0\gg\Omega_0$, (ii)
$\chi_0\approx\Omega_0$, and (iii) $\chi_0\ll\Omega_0$.

In the first case of $\chi_0\gg\Omega_0$, we assume that $\Omega_0$ in
the exponent is negligible; then, a separation of variables is
possible. Imposing the boundary condition of $\Omega_0>0$, a solution
is obtained in the form of
$\Psi_{\alpha,\zeta}=\sin\omega\Omega_0\Psi_\zeta(\chi_0)$, where
$\zeta=\alpha-\kappa\omega^2/4$ and $\Psi_\zeta$ satisfies the
following Liouville-type equation,
\begin{equation}
\left[ \frac\kappa4 \frac{\partial^2}{\partial\chi_0^2} - \zeta -
  4\lambda^2 e^{2\chi_0/\kappa} \right] \Psi_\zeta = 0.  \label{eq:i:chi}
\end{equation}
Fortunately, it can be written as a (modified) Bessel's equation by
defining $\rho=e^{\chi_0/\kappa}$,
\begin{equation}
\left[ \rho^2 \frac{\partial^2}{\partial \rho^2} + \rho 
\frac{\partial}{\partial \rho} - 4\kappa (
  4\lambda^2 \rho^2 + \zeta ) \right] \Psi_\zeta = 0.  \label{eq:chi:Bessel}
\end{equation}
Note that its solution is either a (damping) modified Bessel function
for $\zeta>0$ or a (oscillating) Bessel function for $\zeta<0$. For
positive $\zeta$, the solution is
$\Psi_\zeta=AK_\nu(k\rho)+BI_\nu(k\rho)$, where $A$ and $B$ are
constants, $\nu=2\sqrt{\zeta\kappa}$, and
$k=4\lambda\sqrt{\kappa}$. In the limit of $\rho\to\infty$, the
potential is infinite so that the constant $B$ is set to zero. It
should be pointed out that in the limit of $\rho\to0$, the function
$K_\nu$ seems to diverge; however, this is not the case because it
conflicts with the assumption $\chi_0\gg\Omega_0$. Thus, the finite
solution resulting in this case is
$\Psi_{\alpha}=\sin\omega\Omega_0K_\nu(k\rho)$, which is, in fact, a
quantum wormhole solution~\cite{hp}. For the negative case of
$\zeta\equiv-\xi$, the solution is
$\Psi_\zeta=\tilde{A}J_{\tilde{\nu}}(k\rho)+\tilde{B}N_{\tilde{\nu}}(k\rho)$,
where $\tilde{A}$ and $\tilde{B}$ are constants and
$\tilde{\nu}=2\sqrt{\xi\kappa}$.

In the case of (ii) $\chi_0\approx\Omega_0$,
Eq.~(\ref{eq:WDW:chiOmega}) is written as
\begin{equation}
\left[ \frac\kappa4 \left( \frac{\partial^2}{\partial\chi_0^2} -
  \frac{\partial^2}{\partial\Omega_0^2} \right) - \alpha - 4\lambda^2
  \right] \Psi_\alpha = 0.  \label{eq:ii:chiOmega}
\end{equation}
Then, the solution is
$\Psi_{\alpha}=\sin\omega\Omega_0\Psi_\eta(\chi_0)$, where
$\eta=\kappa\omega^2/4-\alpha-4\lambda^2$ and $\Psi_\eta\sim e^{\pm
  i\eta\chi_0}$. For the case of $\chi_0\ll\Omega_0$, the exponential
term in Eq.~(\ref{eq:WDW:chiOmega}) is negligible, and the solution is
similar to that of case (ii),
$\Psi_{\alpha}=\sin\omega\Omega_0\Psi_\xi(\chi_0)$, where
$\xi=\kappa\omega^2/4-\alpha$ and $\Psi_\xi\sim e^{\pm i\xi\chi_0}$.

We have studied the wave function of the universe in the BPP model by
considering some boundary conditions. The physical significance of the
boundary condition $e^{-2\phi}>0$ is that the curvature singularity in
the semi-classical region does not appear as seen in
Eq.~(\ref{eq:R:scalar}) because the curvature singularity appears  at
$e^{-2\phi}|_{t_a, t_b}=0$. Furthermore, this condition has been
imposed on the wave function of the WD equation similar to the
potential problem in quantum mechanics in the forbidden
region. Therefore, the consistent requirement of boundedness is
directly connected to the avoidance of the singularity. In fact, this
situation appears to be similar to the RST model, where we do not
repeat the same calculations.

\begin{acknowledgments}
W.~Kim is supported by the Science Research Center Program of the
Korea Science and Engineering Foundation through the Center for
Quantum Spacetime (CQUeST) of Sogang University with grant number R11
- 2005 - 021. E.~J.~Son is supported by the Korea Research Foundation
Grant funded by the Korea Government(MOEHRD, Basic Reasearch Promotion
Fund) (KRF-2005-070-C00030). M.~S.~Yoon is supported by the Korea
Research Foundation Grant funded by the Korean Government 
(MOEHRD)(KRF-2005-037-C00017).
\end{acknowledgments}

\end{document}